\documentstyle[multicol,prb,aps,epsf]{revtex}
\begin{document}
\title{Pauli susceptibility of A$_3$C$_{60}$ (A=K, Rb)  }

\author{F. Aryasetiawan$^{(1,2)}$, O. Gunnarsson$^{(1)}$, 
E. Koch$^{(3)}$ and R.M. Martin$^{(3)}$}
\address{${}^{(1)}$Max-Planck-Institut f\"ur Festk\"orperforschung, 
D-70506 Stuttgart, Germany}
\address{${}^{(2)}$Department of Theoretical Physics, University of Lund,
S\"olvegatan 14 A, S-223 62 Lund, Sweden}
\address{${}^{(3)}$Department of Physics, University of Illinois, 
Urbana, Illinois 61801}
\date{\today}
\maketitle
\pacs{71.20.Tx, 75.40C, 71.10.Fd}
\begin{abstract}
The Pauli paramagnetic susceptibility of A$_3$C$_{60}$ (A= K, Rb) 
compounds is calculated.
A lattice quantum Monte Carlo method is applied to 
a multi-band Hubbard model, including the on-site Coulomb interaction $U$.
It is found that the many-body enhancement of the susceptibility is of
the order of a factor of three. This reconciles estimates of the density
of states from the susceptibility with other estimates. The enhancement
is an example of a substantial many-body effect in the doped fullerenes.  

\end{abstract}
\begin{multicols}{2}
The Pauli paramagnetic susceptibility $\chi$ is interesting for several reasons.
Firstly, $\chi$ is enhanced by many-body effects relative to its value $\chi_0$
for noninteracting electrons, and $\chi/\chi_0$ is one measure of the
strength of the many-body effects in the system. In the alkali-doped C$_{60}$
compounds, A$_3$C$_{60}$ (A= K, Rb) the Coulomb interaction $U$
between two electrons on the same molecule is large\cite{Sawatzky}
 compared with the 
width $W$ of the partly filled $t_{1u}$ band, 
with typical estimates $U/W\sim 1.5-2.5$.\cite{C60Mott}   
In view of this large ratio,
one expects very strong many-body effects for these systems. Up to
now, however, there seems to be no unambiguous signature of such 
strong effects.  

Secondly, $\chi_0$ is related to the density of states $N(0)$ at the
Fermi energy, and values of $N(0)$ can be extracted from $\chi$ if
the enhancement $\chi/\chi_0$ is known. $N(0)$ is important for the 
superconductivity and the electron-phonon interaction $\lambda$, since
theoretical calculations give $\lambda/N(0)$, while some    experimental
(e.g., neutron\cite{Prassides} and Raman scattering\cite{Winter})
estimates give $\lambda N(0)$ and others (e.g., photoemission from
C$_{60}^-$ molecules\cite{PES}) give $\lambda/N(0)$. 
The estimate of $N(0)$ is therefore crucial for obtaining values of  $\lambda$
and for our understanding of the superconductivity.
Typically, for K$_3$C$_{60}$
band structure calculations give $N(0)\sim 6-9$ 
states/(eV-spin).\cite{Pickett,Martins,Huang,Satpathy,Novikov,WAntropov}
Estimates based on the specific heat and the NMR relaxation rate give 
$N(0)\sim 5-6$ states/(eV-spin)\cite{Meingast} 
and $N(0)\sim 7.2$ states/(eV-spin),\cite{Antropov} respectively.  
On the other hand,   
much larger values $N(0)\sim 10-16$ 
states/(eV-spin) for K$_3$C$_{60}$ are deduced from 
the susceptibility,\cite{Ramirez,Wong,Tanigaki,Wang} if          
many-body effects are neglected.
A substantial many-body enhancement (factor 2-3)
for the susceptibility,\cite{Ramirez} could essentially reconcile 
these rather  different estimates. On the other hand, density functional
calculations in the local density approximation (LDA) find that the
enhancement is only about a factor of 1.3-1.4.\cite{Antropov}
We note that $N(0)$ here refers to densities of states defined in 
somewhat different ways depending on the experiment, as discussed below.

We have used a lattice quantum Monte Carlo method\cite{tenHaaf,C60Mott}
 for calculating
the Pauli susceptibility for a multi-band Hubbard model of the system. For a
small system with four C$_{60}$ molecules, we demonstrate that this method
gives an accurate enhancement of the susceptibility. For realistic
values of the parameters, the susceptibility is enhanced by about a 
factor of three, which essentially reconciles estimates of $N(0)$
based on the susceptibility with other estimates.

In the presence of a small external magnetic field $\mathcal{H}$, 
the energy of the system can be written as
\begin{equation}\label{eq:1}
E({\mathcal{M}})=E_0({\mathcal{M}})-{\mathcal{M}}{\mathcal{H}}
\approx E_{00} +{1\over 2}\alpha {\mathcal{M}}^2 -{\mathcal{M}\mathcal{H}},
\end{equation}
where ${\mathcal{M}}\equiv -\mu_B(N_{\uparrow}-N_{\downarrow})$ is the
magnetic moment of the system, with $\mu_B$ being the Bohr magneton and
$N_{\sigma}$ the number of electrons with spin $\sigma$. $E_0({\mathcal{M}})$
is the energy of the system with a moment $\mathcal{M}$ in the absence
of an external field.
Minimizing the energy with respect to $\mathcal{M}$, we obtain
the susceptibility
\begin{equation}\label{eq:2}
\chi\equiv {\mathcal{M}\over \mathcal{H}}={1 \over \alpha}.
\end{equation}
In the following we therefore calculate $E_0({\mathcal{M}})$ for 
the interacting and noninteracting ($U=0$) systems, from which we
obtain the many-body enhancement $\chi/\chi_0$.

We use a multi-band Hubbard model of the A$_3$C$_{60}$ compounds 
\begin{eqnarray}\label{eq:3}
H=&&\sum_{i\sigma}\sum_{m=1}^3\varepsilon_{t_{1u}}n_{i\sigma m}+\sum_{<ij>\sigma
mm'}
t_{ijmm'}\psi^{\dagger}_{i\sigma m} \psi_ {j\sigma m'} \nonumber \\
+ &&
U\sum_i\sum_{\sigma m < \sigma'm'}n_{i\sigma m}n_{i\sigma'm'},
\end{eqnarray}
where the first term describes the three-fold degenerate $t_{1u}$
states on the sites $i$ and with orbital ($m$) and spin ($\sigma$) indices.
The second term describes  the hopping between the sites,  
and the third term describes the Coulomb on-site
interaction. Multiplet effects and the electron-phonon interaction 
have been neglected. In A$_3$C$_{60}$, the C$_{60}$ molecules are
 preferentially in one of two possible orientations in an essentially
random way.\cite{Stephens} We take this into account by having a large 
cell, where each molecule takes one of the two preferred 
orientations in a random way,
 and the hopping matrix
elements between two molecules take into account the orientations
of these two molecules.\cite{Orientation,MazinAF}

To calculate the energy of the model in Eq.~(\ref{eq:3}), we use a 
$T=0$ projection lattice Monte Carlo method,
 introduced by ten Haaf {\it et al}.\cite{tenHaaf}
In this method a trial function is constructed from a Slater determinant,
using a Gutzwiller Ansatz.\cite{Gutzwiller} An approximate ground-state
is then projected out in a diffusion Monte Carlo (DMC) approach, using a   
``fixed node'' approximation. This method has been used to study 
the condition for a Mott-Hubbard transition in A$_3$C$_{60}$.\cite{C60Mott}

To test the accuracy of the DMC approach, we have first applied the method
to a cluster of four C$_{60}$ molecules. This cluster is so small that 
we can also obtain the exact solution for the model in Eq.~(\ref{eq:3}),
using exact diagonalization.
We then calculate the coefficient $\alpha$ in Eq.~(\ref{eq:1}) by
considering the energy for $N_\uparrow-N_\downarrow=$ 0 and 2. The
DMC and exact results are compared in Table \ref{tableI}. We can see 
that the DMC method is quite accurate in this case, and if a similar 
accuracy is obtained for larger systems, it is quite sufficient.

\noindent
\begin{minipage}{3.375in}
\begin{table}[h]  
\caption[]{The enhancement of the susceptibility for a model with four 
C$_{60}$ molecules according to diffusion Monte Carlo (DMC)
and exact calculations as a function of the Coulomb energy $U$. The band width
is $W=0.58$ eV.}                             
\begin{tabular}{ccc}
$U$  & \multicolumn{2}{c}{$\chi/\chi_0$} \\
\tableline
       &   DMC     & Exact   \\
\tableline
1.0    &  1.89     & 1.93    \\
1.25   &  2.20     & 2.26    \\
1.50   &  2.63     & 2.69    \\
\end{tabular}\label{tableI}
\end{table}
\end{minipage}

The enhancement of the susceptibility is sensitive to the density 
of states (DOS) close to the Fermi energy $E_F$.
For small and intermediate size clusters of C$_{60}$ molecules,
the DOS depends on the orientations of the molecules, while for large
clusters the DOS rapidly converges. Since we can only treat intermediate
size clusters ($\sim (32-64)$ molecules) in DMC,
 we have therefore chosen orientations which 
in a one-particle approximation give 
similar DOS close to $E_F$ as for very large clusters.  

\begin{figure}[bt]
\unitlength1cm
\begin{minipage}[t]{8.5cm}
\centerline{\epsfxsize=3.375in \epsffile{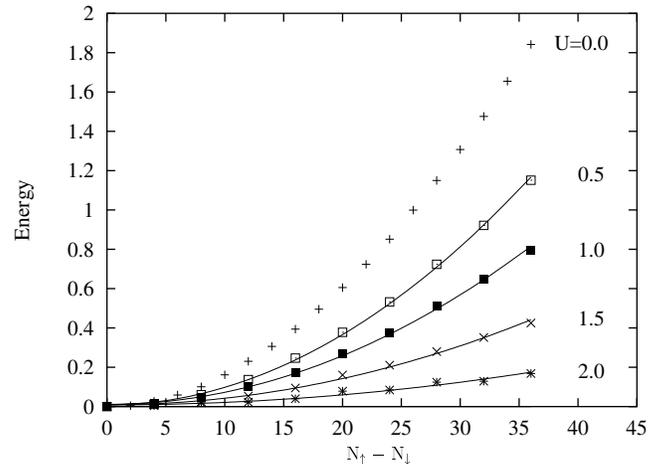}}
\caption[]{\label{fig1}The energy as a function of $N_{\uparrow}-N_{\downarrow}
\sim {\mathcal{M}}$ relative to the energy for $\mathcal{M}$=0.
Second order curves have been fitted to the energies. Results for 
$U=$ 0, 0.5, 1.0, 1.5 and 2.0 eV are shown.
The calculations are for 32 molecules and the energy is in eV.} 
\end{minipage}
\hfill
\end{figure}

\begin{figure}[bt]
\unitlength1cm
\begin{minipage}[t]{8.5cm}
\centerline{\epsfxsize=3.375in \epsffile{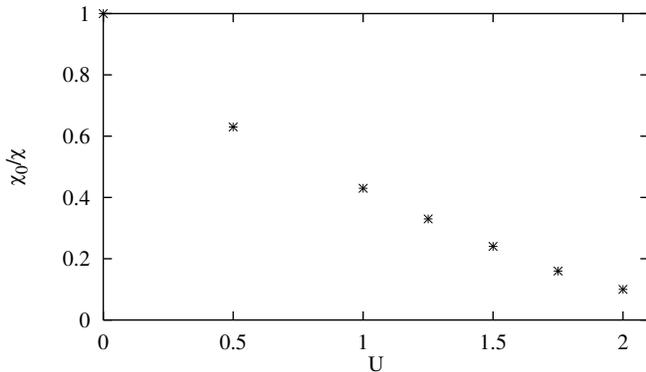}}
\caption[]{\label{fig2}The inverse enhancement $\chi_0/\chi$ of the 
susceptibility for a cluster with 32 molecules as a function of the Coulomb 
interaction $U$ (in eV). The band width is 0.66 eV.}                                        
\end{minipage}
\hfill
\end{figure}

In Fig. \ref{fig1} we show results for the total energy as a 
function of ${\mathcal{M}}\sim N_{\uparrow}-N_{\downarrow}$ for
different values of $U$. 
The results can be rather well fitted by parabolas, although the 
precise parameters of the parabolas have a certain dependence 
on the range of $\mathcal{M}$ considered. From these slopes
we can immediately deduce values of the enhancement $\chi/\chi_0$.
In Fig. \ref{fig2} the inverse of the enhancement $\chi_0/\chi$ is 
shown. It is immediately clear that the enhancement grows with $U$
and that $\chi$ would diverge for $U$ a bit larger than 2 eV, if no other 
transition (e.g., antiferromagnetic) happened before.
Estimates of $U$ are typically in the range 1-1.5 eV, giving an 
enhancement of the susceptibility by about a factor of three.
Qualitatively similar results have been obtained for a   Hubbard
model without orbital degeneracy and in the limit of infinite
dimensions.\cite{Georges}

\noindent

Within the Hartree-Fock approximation, the susceptibility behaves as
\begin{equation}\label{eq:4}
{\chi_0 \over \chi}\sim 1-{N(0)   \over 3}U,
\end{equation}
where the factor three comes from the three-fold degeneracy
and $N(0)$ is the density of states per spin, which here is about 
5.5 states/(eV-spin). Here we have assumed that the three orbitals are 
equivalent.
The DMC results show a similar behavior, but with a 
prefactor in front of $U$, which is between a factor of four  and five 
times smaller.
This large change in the prefactor illustrates the importance of correlation
in these systems.

To deduce $N(0)$ we need to know the Pauli (paramagnetic)  susceptibility.
 Measurements using a SQUID may also contain a diamagnetic contribution,
while EPR measurements do not. In Table \ref{tableII} we show various
experimental results converted to $N(0)$. From the SQUID results,
diamagnetic contributions estimated by the respective authors have been 
subtracted, but the many-body enhancement has {\em not} been considered.  
We can see that the results range between 10 and 16 states/(eV-spin). 
If we consider a many-body enhancement of a factor of three, as deduced above,
these results would be reduced to about $N^{susc}(0)\sim4-5$ 
states/(eV-spin).

\noindent

Up to now we have not fully considered differences in the  
definition of the density of states $N(0)$ deduced from different 
sources. Calculations of the electron-phonon
interaction $\lambda$ are usually based on band structure calculations,
giving a density of states $N_{0}(0)$ for noninteracting electrons. 
The electron-electron interaction modifies $N_0(0)$ to its interacting
value ${\mathcal{N}}(0)$ (density of quasi-particle states), which 
should be used in the calculation of $\lambda$.\cite{Scalapino}
This density of states enters in the specific heat\cite{Pines}
\begin{equation}\label{eq:4a}
{C_v \over T}\sim {\mathcal{N}}(0)(1+\lambda),
\end{equation}
where the last factor is due to the electron-phonon interaction,
assuming that the band width is large compared with the phonon 
frequency.\cite{Grimvall}
In the same approximation,
the susceptibility can be written as\cite{Pines}
\begin{equation}\label{eq:4b}
\chi \sim {{\mathcal{N}}(0)\over 1+F_0^a},
\end{equation}
where $F_0^a$ is a Landau parameter and $1/(1+F_0^a)$ can be considered
as a Stoner enhancement.
Our calculation $\chi/chi_0$ includes both the factors 
${\mathcal N}(0)/N_0(0)$ 
and $(1+F_0^a)$.  Within the present Monto
Carlo technique it is not possible to calculate the specific heat,
and we can therefore not separate the two contributions. 
The value of $N^{susc}(0)$ deduced from the susceptibility,
after dividing out many-body effects,  should therefore
primarily be compared with the results obtained from
band structure calculations. The band structure calculations have 
been performed for orientationally ordered systems and give
$N_0^{ord}\sim 6-9$ states/(eV-spin). Sine the real systems have
orientational disorder, we estimate the corresponding density of states
by solving the Hamiltonian (\ref{eq:3}) for $U=0$ and orientational
disorder. Comparision with calculations for ordered systems suggests
a reduction of $N(0)$ by about $15-20 \%$ to $N_0^{disord}(0)\sim 5-7$  
states/(eV-spin). This is in rather good agreement with 
$N^{susc}(0)\sim 4-5$ states/(eV-spin).

It is interesting    that the estimate of ${\mathcal N}(0)\sim 5-6$ states
/(eV-spin) based on the specific 
heat\cite{Meingast} is comparable to the noninteracting result
$N_0^{disord}(0)\sim 5-7$ states/(eV-spin). In contrast to
what has been found for the nondegenerate Hubbard model in 
infinite dimensions,\cite{Georges} this surprising observation suggests that 
the enhancement of the density of states is small in 
the A$_3$C$_{60}$ compounds or   that there may even be a reduction.  
If this is indeed the case, then the enhancement of the susceptibility would 
be a Stoner enhancement, and the $N_0^{susc}(0)$ obtained from the susceptibility,
after dividing out the many-body enhancement, could be compared with other
experimental estimates of $N(0)$. We then find that these experimental 
estimates are essentially brought in line with each other
and with the band structure estimates, giving
$N(0)\sim 5-7$ states/(eV-spin) for K$_3$C$_{60}$.

In our model and in the calculations of the susceptibility, we have 
neglected multiplet effects and the electron-phonon interaction, 
which are now discussed.
The electron-phonon interaction leads to 
an increase of the density of states at the Fermi energy, 
due to the reduced dispersion of states within roughly a phonon
energy of the Fermi energy.
This does not influence the susceptibility, if the phonon energies are
small compared with the electronic energies.\cite{Prange}
Although this assumption    may not be entirely satisfied for A$_3$C$_{60}$ 
and interesting effects may result from the finite band width,\cite{Sasha}
we here neglect the  effects of the electron-phonon interaction
on the density of states.
Instead we focus on how the electron-phonon interaction influences 
the moment formation on the C$_{60}$ molecules.

\noindent
\begin{minipage}{3.375in}
\begin{table}[h]  
\caption[]{The density of states $N(0)$ (per eV and spin) for
 K$_3$C$_{60}$ 
as deduced from susceptibility measurements. The results have
{\sl not} been corrected for the Stoner enhancement, which 
would lead to  reduced    estimates of $N(0)$.}
\begin{tabular}{cll}
$N(0)$ (K$_3$C$_{60}$)   &    Method & Reference \\
\tableline
  14     &      SQUID & Ramirez {\it et al.}\cite{Ramirez} \\
  16     &            SQUID & Wong    {\it et al.}\cite{Wong}   \\
  11     &            EPR& Wong    {\it et al.}\cite{Wong}      \\
  15     &      EPR & Tanigaki {\it et al.}\cite{Tanigaki}      \\
  10     &            EPR& Wang    {\it et al.}\cite{Wang}             \\
\end{tabular}
\label{tableII}
\end{table}
\end{minipage}

If the multiplets are neglected but the electron-phonon interaction
(Jahn-Teller effect) is considered, the lowest spin 1/2 state is favored
over the spin 3/2 state, according to calculations for the lowest 
state of each multiplicity for a free molecule.\cite{Auerbach} The energy 
lowering of the spin 1/2 relative to 
the spin 3/2 state may be as large as  0.3 eV.\cite{c60jt} 
This suppresses the formation of moments and probably tends to reduce
the susceptibility. On the other hand,
 the multiplet effects should favor the formation of moments
on the C$_{60}$ molecules, by giving preference to states with the spin
3/2.  

The multiplet effects lead to   five spin 1/2 states with the energy 
$3K$ and three spin 1/2 states with the energy $5K$ relative to the
spin 3/2 state. Here $K$ is the exchange integral between two $t_{1u}$
orbitals, and we have assumed that the Coulomb integral ($U_{xx}$)  between 
equal orbitals is $2K$ larger than the one ($U_{xy}$) between different 
orbitals. To estimate $K$,
we have used a simple model, where the Coulomb integral between 
two $2p$-charge distributions on two carbon atoms goes as $e^2/R$,
where $R$ is the separation between the two atoms.\cite{Trieste}
 The on-site interaction was assumed to be 15 eV.   
Without screening we find that $K=0.12$ eV. This probably overestimates
the multiplet effects. An alternative estimate is 
obtained by using RPA screening of the Coulomb interaction. We then
find $K=0.030$ eV. A similar result (0.024 eV) was also found by 
Joubert, using a density functional approach.\cite{Joubert}

From the numbers above it follows that there should be a partial cancellation
between electron-phonon and multiplet effects. 
Depending on which numbers are used, either effect could be argued to be larger.
If the electron-phonon effects win, this may lead                 
to a somewhat smaller enhancement of the susceptibility
than was found above (Fig. \ref{fig2}).

We have calculated the Pauli susceptibility of the doped fullerenes
A$_3$C$_{60}$ (A= K, Rb).
The enhancement is of the order of a factor of three,
which allows us to reconcile the estimates of the density of states
from the susceptibility with other estimates.
This suggests that for K$_3$C$_{60}$ $N(0)\sim 5-7$ states/(eV-spin).
This value is only slightly smaller than a value ($N(0)=7.2$) used recently
to provide support for an electron-phonon mechanism driving
the superconductivity in K$_3$C$_{60}$,\cite{PES} but substantially
smaller than some values used in early theoretical discussions.
The susceptibility   enhancement is appreciably   larger than
the one (factor 1.3-1.4) found in the LDA, and it is one of the
first explicit examples of important many-body effects, expected to be
found in these systems. Comparison with Hartree-Fock calculations shows,
however, that the enhancement is about four to five times smaller than
the HF result, illustrating the importance of correlation effects.

\end{multicols}

\begin{thebibliography}{*}

\bibitem{Sawatzky}
R.W. Lof,  M.A. van Veenendaal, B. Koopmans, H.T. Jonkman, and G.A.
Sawatzky, 1992, Phys. Rev. Lett. {\bf 68}, 3924.

\bibitem{C60Mott}O. Gunnarsson, E. Koch, and R.M. Martin, Phys. Rev. B
{\bf 54}, R11026 (1996).

\bibitem{Prassides}
K. Prassides, C. Christides, M.J. Rosseinsky, J. Tomkinson,
D.W. Murphy, and R.C. Haddon, Europhys. Lett. {\bf 19}, 629 (1992).


\bibitem{Winter}   
Winter, J. and H. Kuzmany, 1996, Phys. Rev. B {\bf 53}, 655.

\bibitem{PES}O. Gunnarsson, H. Handschuh, P.S. Bechthold,
B. Kessler, G. Gantef\"or, and W. Eberhardt, Phys. Rev. Lett.
{\bf 74}, 1875 (1995).

\bibitem{Pickett}
S.C.~Erwin and W.E.~Pickett, Science {\bf 254}, 842 (1991).

\bibitem{Martins}
N.~Troullier and J.L.~Martins, Phys.~Rev.~B {\bf 46}, 1766 (1992).
 
\bibitem{Huang}
M.Z. Huang, Y.-N. Xu and W.Y. Ching, Phys. Rev. B {\bf 46}, 6572 (1992).

\bibitem{Satpathy}
S. Satpathy,  V.P. Antropov, O.K. Andersen, O. Jepsen, O. Gunnarsson,
and A.I. Liechtenstein, Phys. Rev. B {\bf 46}, 1773 (1992).

\bibitem{Novikov}
D.L. Novikov, V.A. Gubanov and A.J. Freeman, Physica C {\bf 191}, 399 (1992).

\bibitem{WAntropov}
V.P.~Antropov {\it et al.} (unpublished).


\bibitem{Meingast}
G.J. Burkhart, and C. Meingast, Phys. Rev. B {\bf 54}, R6865 (1996). 

\bibitem{Antropov}V.P. Antropov, I.I. Mazin, O.K. Andersen, A.I. 
Liechtenstein, and O. Jepsen, Phys. Rev. B {\bf 47}, 12 373 (1993).


\bibitem{Ramirez}
A.P. Ramirez, M.J. Rosseinsky, D.W. Murphy, and R.C. Haddon, Phys. Rev. Lett.
{\bf 69}, 1687 (1992).

\bibitem{Wong}
W.H. Wong, M.E. Hanson, W.G. Clark, G. Gr\"uner, J.D. Thompson,
R.L. Whetten, S.-M. Huang, R.B. Kaner, F. Diederich, P. Petit,
J.-J. Andre and K. Holczer, Europhys. Lett. {\bf 18}, 79 (1992).

\bibitem{Tanigaki}
K. Tanigaki, M. Kosaka, T. Manako, Y. Kubo, I. Hirosawa, K. Uchida,
and K. Prassides, Chem. Phys. Lett. {\bf 240}, 627 (1995).

\bibitem{Wang}
D. Wang, Solid State Commun. {\bf 94}, 767 (1995).

\bibitem{tenHaaf}   
F.B. ten Haaf, H.J.M. van Bemmel,
J.M.J. van Leeuwen, W. van Saarloos, and D.M. Ceperley, Phys. Rev. B
{\bf 51}, 353 (1995); H.J.M. van Bemmel, D.F.B. van Haaf, W. van Saarloos,
J.M.J. van Leeuwen, and G. An, Phys. Rev. Lett. {\bf 72}, 2442 (1994).

\bibitem{Stephens} P.W. Stephens, L. Mihaly, P.L. Lee, R.L. Whetten, 
S.-M. Huang, R. Kaner, F. Deiderich, and K. Holczer, 
1991, Nature {\bf 351}, 632.

\bibitem{Orientation}O. Gunnarsson, S. Satpathy, O. Jepsen, and O.K. Andersen,
Phys. Rev. Lett. {\bf 67}, 3002 (1991).

\bibitem{MazinAF} I.I. Mazin, A.I. Liechtenstein, O. Gunnarsson, O.K.
Andersen, V.P. Antropov, and S.E. Burkov, Phys. Rev. Lett.
{\bf 26}, 4142 (1993).


\bibitem{Gutzwiller}
M.C.~Gutzwiller et al., Phys.~Rev.~{\bf 137}, A1726 (1965).

\bibitem{Scalapino}
D.J. Scalapino, in {\it Superconductivity}, edited by R.D. Parks, 
Dekker (New York, 1969), p. 449. 

\bibitem{Pines}
D. Pines and P. Noziere, {\it The theory of quantum liquids},
Benjamin (New York, 1966).

\bibitem{Grimvall}  G. Grimvall, {\sl The electron-phonon interaction in
metals}, North-Holland, (1981), p. 212.

\bibitem{Georges}
 A.~Georges, G. Kottliar, W. Krauth, and M.J. Rozenberg, 
Rev.~Mod.~Phys.~{\bf 68}, 13 (1996).
 

\bibitem{Prange}
R.E. Prange and A. Sachs, Phys. Rev. {\bf 158}, 672 (1967).

\bibitem{Sasha}
A.I. Liechtenstein, O. Gunnarsson, M. Knupfer, J. Fink, 
and J.F. Armbruster, J. Phys.: Cond. Matter {\bf 8}, 4001 (1996).

\bibitem{Auerbach}A. Auerbach, N. Manini and E. Tosatti,
Phys. Rev. B {\bf 49}, 12998 (1994).
N. Manini, E. Tosatti and A. Auerbach, Phys. Rev. {\bf 49}, 13008 (1994).
\bibitem{c60jt}
O. Gunnarsson, Phys. Rev. B {\bf 51}, 3493 (1995).
\bibitem{Trieste}Gunnarsson, O., D. Rainer, and G. Zwicknagl, 
1992, Int. J. Mod. Phys. B {\bf 6}, 3993.

\bibitem{Joubert}
D.P. Joubert, J. Phys.: Condens. Matter {\bf 5}, 8047 (1993).


\end{thebibliography}
\end{document}